\documentclass[a4paper,10pt]{article}
\usepackage{graphicx}
\begin{document}

\title{\bf Expressing Algorithms As Concise As Possible via Computability Logic}
\author{Keehang Kwon\\
\sl \small Faculty of Computer Engineering, DongA  University\\
\sl \small 840 Hadan Saha, 604-714 Busan, Korea\\
\small  khkwon@dau.ac.kr}
\date{}
\maketitle

%% <local definitions here>

\makeatletter
\long\def\@makemyfntext#1{$^{\rm *}\ $ #1}

\long\def\@myfootnotetext#1{\insert\footins{\footnotesize
    \interlinepenalty\interfootnotelinepenalty 
    \splittopskip\footnotesep
    \splitmaxdepth \dp\strutbox \floatingpenalty \@MM
    \hsize\columnwidth \@parboxrestore
   \edef\@currentlabel{\csname p@footnote\endcsname\@thefnmark}\@makemyfntext
    {\rule{\z@}{\footnotesep}\ignorespaces
      #1\strut}}}

\def\myfootnotetext{\@ifnextchar
     [{\@xfootnotenext}{\xdef\@thefnmark{\thempfn}\@myfootnotetext}}
\makeatother
%\setstretch{2}

\newcommand{\gneg}{\neg} % negation

\newcommand{\mlc}{\wedge} % parallel conjunction

\newcommand{\mld}{\vee} % parallel disjunction

\newcommand{\mli}{\rightarrow} % basic reduction

\newcommand{\mla}{\mbox{{\Large $\wedge$}}} % parallel universal quantifier

\newcommand{\mle}{\mbox{{\Large $\vee$}}} % parallel existential quantifier

\newcommand{\pst}{\mbox{\raisebox{-0.01cm}{\scriptsize $\wedge$}\hspace{-4pt}\raisebox{0.16cm}{\tiny $\mid$}\hspace{2pt}}} % parallel recurrence

\newcommand{\pcost}{\mbox{\raisebox{0.12cm}{\scriptsize $\vee$}\hspace{-4pt}\raisebox{0.02cm}{\tiny $\mid$}\hspace{2pt}}} % parallel corecurrence

\newcommand{\pintimpl}{\mbox{\hspace{2pt}\raisebox{0.033cm}{\tiny $>$}\hspace{-0.18cm} \raisebox{-0.043cm}{\large --}\hspace{2pt}}} % parallel-recurrence-based reduction

\newcommand{\cla}{\mbox{\large $\forall$}} % blind universal quantifier

\newcommand{\cle}{\mbox{\large $\exists$}} % blind existential quantifier

\newcommand{\adc}{\sqcap} % choice conjunction

\newcommand{\add}{\sqcup} % choice disjunction

\newcommand{\adi}{\sqsupset} % choice implication

\newcommand{\ada}{\mbox{\Large $\sqcap$}} % choice universal quantifier

\newcommand{\ade}{\mbox{\Large $\sqcup$}} % choice existential quantifier

\newcommand{\st}{\mbox{\raisebox{-0.05cm}{$\circ$}\hspace{-0.13cm}\raisebox{0.16cm}{\tiny $\mid$}\hspace{2pt}}} % branching recurrence

\newcommand{\cost}{\mbox{\raisebox{0.12cm}{$\circ$}\hspace{-0.13cm}\raisebox{0.02cm}{\tiny $\mid$}\hspace{2pt}}} % branching corecurrence

\newcommand{\intimpl}{\mbox{\hspace{2pt}$\circ$\hspace{-0.14cm} \raisebox{-0.043cm}{\Large --}\hspace{2pt}}} % branching-recurrence-based reduction

\newcommand{\sqc}{\mbox{\small \raisebox{0.0cm}{$\bigtriangleup$}}} % sequential conjunction

\newcommand{\sqd}{\mbox{\small \raisebox{0.049cm}{$\bigtriangledown$}}} % sequential disjunction

\newcommand{\sqa}{\mbox{\large \raisebox{0.0cm}{$\bigtriangleup$}}} % sequential universal quantifier

\newcommand{\sqe}{\mbox{\large \raisebox{0.07cm}{$\bigtriangledown$}}} % sequential existential quantifier

\newcommand{\sst}{\mbox{\raisebox{-0.07cm}{\scriptsize $-$}\hspace{-0.2cm}$\pst$}} % sequential recurrence

\newcommand{\scost}{\mbox{\raisebox{0.20cm}{\scriptsize $-$}\hspace{-0.2cm}$\pcost$}} % sequential corecurrence

\newcommand{\sintimpl}{\mbox{\hspace{2pt}\raisebox{0.033cm}{\tiny $ | \hspace{-4pt} >$}\hspace{-0.14cm} \raisebox{-0.039cm}{\large --}\hspace{2pt}}} % sequential-recurrence-based reduction

\newcommand{\cst}{{\mbox{\raisebox{-0.05cm}{$\circ$}\hspace{-0.13cm}\raisebox{0.16cm}{\tiny $\mid$}\hspace{1pt}}}^{\aleph_0}} % countable recurrence

\newcommand{\ccost}{{\mbox{\raisebox{0.12cm}{$\circ$}\hspace{-0.13cm}\raisebox{0.02cm}{\tiny $\mid$}\hspace{1pt}}}^{\aleph_0}} % countable corecurrence

\newcommand{\cintimpl}{{\mbox{\hspace{2pt}$\circ$\hspace{-0.14cm} \raisebox{-0.043cm}{\Large --}\hspace{1pt}}}^{\aleph_0}} % countable-recurrence-based reduction

\newcommand{\fintimpl}{\mbox{\hspace{2pt}$\bullet$\hspace{-0.14cm} \raisebox{-0.058cm}{\Large --}\hspace{-6pt}\raisebox{0.008cm}{\scriptsize $\wr$}\hspace{-1pt}\raisebox{0.008cm}{\scriptsize $\wr$}\hspace{4pt}}} % dfb-reduction

\newcommand{\sfbr}{\mbox{\hspace{2pt}$\bullet$\hspace{-0.14cm} \raisebox{-0.058cm}{\Large --}\hspace{-6pt}\hspace{1pt}\raisebox{0.008cm}{\scriptsize $\wr$}\hspace{4pt}}} % sfb-reduction

\newcommand{\fbr}{\mbox{\hspace{2pt}$\bullet$\hspace{-0.14cm} \raisebox{-0.058cm}{\Large --}\hspace{2pt}}} % fb-reduction

　

\newcommand{\tgd}{\mbox{\hspace{2pt}$\vee$\hspace{-1.29mm}\raisebox{0.1mm}{\rule{0.13mm}{2mm}}\hspace{5pt}}} % toggling disjunction

\newcommand{\tgc}{\mbox{\hspace{2pt}$\wedge$\hspace{-1.29mm}\raisebox{0.02mm}{\rule{0.13mm}{2mm}}\hspace{5pt}}} % toggling conjunction

\newcommand{\tge}{\hspace{1pt}\mbox{\Large $\vee$\hspace{-1.84mm}\raisebox{0.1mm}{\rule{0.13mm}{3.0mm}}\hspace{6pt}}} % toggling existential quantifier

\newcommand{\tga}{\mbox{\hspace{1pt}\Large $\wedge$\hspace{-1.84mm}\raisebox{0.02mm}{\rule{0.13mm}{3.0mm}}\hspace{6pt}}} % toggling universal quantifier

\newcommand{\tgpst}{\mbox{\raisebox{-0.01cm}{\scriptsize $\wedge$}\hspace{-4pt}\raisebox{0.06cm}{\small $\mid$}\hspace{2pt}}} % toggling recurrence

\newcommand{\tgpcost}{\mbox{\raisebox{0.12cm}{\scriptsize $\vee$}\hspace{-3.8pt}\raisebox{0.04cm}{\small $\mid$}\hspace{2pt}}} % toggling corecurrence

\newcommand{\tgst}{\mbox{\raisebox{-0.05cm}{$\circ$}\hspace{-0.12cm}\raisebox{0.05cm}{\small $\mid$}\hspace{2pt}}} % toggling-branching recurrence

\newcommand{\tgcost}{\mbox{\raisebox{0.12cm}{$\circ$}\hspace{-0.12cm}\raisebox{0.04cm}{\small $\mid$}\hspace{2pt}}} % toggling-branching corecurrence

\newcommand{\tgpi}{\mbox{\hspace{2pt}\raisebox{0.033cm}{\tiny $>$}\hspace{-0.28cm} \raisebox{-2.3pt}{\LARGE --}\hspace{2pt}}} % toggling-recurrence-based implication

\newcommand{\tgbi}{\mbox{\hspace{2pt}$\circ$\hspace{-0.26cm} \raisebox{-2.3pt}{\LARGE --}\hspace{2pt}}} % toggling-branching-recurrence-based implication
　
\newenvironment{describe}{\begin{list}{}{\setlength\leftmargin{80pt}}\setlength\labelsep{10pt}\setlength\labelwidth{70pt}}{\end{list}}

\newenvironment{flag}{\begin{list}{\makebox[20pt]{\hss$\circ$\enspace}}
                                  {\labelwidth 20pt}}{\end{list}}

%% js \newtheorem{proposition}{Proposition}

%% js \newenvironment{proof}
     %% js {\begin{trivlist}\item[]{\bf Proof. }}%
     %% js {\\* \hspace*{\fill} $\Box$\end{trivlist}}

\newenvironment{numberedlist}
{\begin{list}{\makebox[20pt]{\hss(\arabic{itemno})\enspace}}
             {\usecounter{itemno}\labelwidth 20pt}}{\end{list}}

\newenvironment{alphabetlist}
{\begin{list}{\makebox[20pt]{\hss(\alph{itemno1})\enspace}}
             {\usecounter{itemno1}\labelwidth 20pt}}{\end{list}}

\newenvironment{romanlist}
{\begin{list}{\makebox[20pt]{\hss(\roman{itemno2})\enspace}}
             {\usecounter{itemno2}\labelwidth 20pt}}{\end{list}}

\newcounter{itemno}

\newcounter{itemno1}

\newcounter{itemno2}
\newcounter{lemma}
\newcounter{exno}

\newcounter{defno}

%\newcounter{exno}[section]

%\newcounter{defno}[section]

%\newtheorem{defn}{Definition}[section]

%\newtheorem{ex}[defn]{Example}

%% js \newtheorem{lemma}{Lemma}

%% js \newtheorem{theorem}[lemma]{Theorem}

\newenvironment{defn}{\refstepcounter{defno}\medskip \noindent {\bf
Definition \thedefno.\ }}{\medskip}

\newenvironment{ex}{\refstepcounter{exno}\medskip \noindent {\bf
Example \theexno.\ }}{\medskip}

\newenvironment{millerexample}{
 \begingroup \begin{tabbing} \hspace{2em}\= \hspace{5em}\= \hspace{5em}\=
\hspace{5em}\= \kill}{
 \end{tabbing}\endgroup}

\newenvironment{wideexample}{
 \begingroup \begin{tabbing} \hspace{2em}\= \hspace{10em}\= \hspace{10em}\=
\hspace{10em}\= \kill}{
 \end{tabbing}\endgroup}

\newcommand{\sep}{\;\vert\;}

\newcommand{\ra}{\rightarrow}
\newcommand{\app}{\ }
\newcommand{\appt}{\ }
\newcommand{\tup}[1]{\langle\nobreak#1\nobreak\rangle}

\newcommand{\hu}{{\cal H}^+}
\newcommand{\Free}{{\cal F}}
\newcommand{\oprove}{\vdash\kern-.6em\lower.7ex\hbox{$\scriptstyle O$}\,}
\newcommand{\true}{\top}

\newcommand{\Dscr}{{\cal D}}
\newcommand{\Pscr}{{\cal P}}
\newcommand{\Gscr}{{\cal G}}
\newcommand{\Fscr}{{\cal F}}
\newcommand{\Vscr}{{\cal V}}
\newcommand{\Uscr}{{\cal U}}
\newcommand{\pderivation}{{\cal P}\kern -.1em\hbox{\rm -derivation}}
\newcommand{\pderivationl}{{\cal P}\kern -.1em\hbox{\em -derivation}}
\newcommand{\pderivable}{{\cal P}\kern -.1em\hbox{\rm -derivable}}
\newcommand{\pderivablel}{{\cal P}\kern -.1em\hbox{\em -derivable}}
\newcommand{\pderivations}{{\cal P}\kern -.1em\hbox{\rm -derivations}}
\newcommand{\pderivability}{{\cal P}\kern -.1em\hbox{\rm -derivability}}
\newcommand{\eqm}[1]{=_{\scriptscriptstyle #1}}
\newcommand\subsl{\preceq}
\newcommand{\fnrestr}{\uparrow}

\newcommand{\match}{{\rm MATCH}}
\newcommand{\triv}{{\rm TRIV}}
\newcommand{\imit}{{\rm IMIT}}
\newcommand{\proj}{{\rm PROJ}}
\newcommand{\simpl}{{\rm SIMPL}}
\newcommand{\failed}{{\bf F}}

\newcommand{\Dsiginst}[1]{{[#1]_\Sigma}}
\newcommand{\Psiginst}[1]{{[#1]_\Sigma}}
\newcommand{\lnorm}{{\lambda}norm}
\newcommand{\seq}[2]{#1 \supset #2}
\newcommand{\dseq}[2]{#1_1,\ldots,#1_{#2}}

\newcommand{\all}{\forall}
\newcommand{\some}{\exists}
\newcommand{\lambdax}[1]{\lambda #1\,}
\newcommand{\somex}[1]{\some#1\,}
\newcommand\allx[1]{\all#1\,}

\newcommand{\subs}[3]{[#1/#2]#3}
\newcommand{\rep}[3]{S^{#2}_{#1}{#3}}
\newcommand{\ie}{{\em i.e.}}
\newcommand{\eg}{{\em e.g.}}

% These are the annotations used with inference figures
\newcommand{\lbotr}{$\bot$-R}
\newcommand{\ldbotr}{\bot\mbox{\rm -R}}
\newcommand{\landl}{$\land$-L}
\newcommand{\ldandl}{\land\mbox{\rm -L}}
\newcommand{\landr}{$\land$-R}
\newcommand{\ldandr}{\land\mbox{\rm -R}}
\newcommand{\lorl}{$\lor$-L}
\newcommand{\ldorl}{\lor\mbox{\rm -L}}
\newcommand{\lorr}{$\lor$-R}
\newcommand{\ldorr}{\lor\mbox{\rm -R}}
\newcommand{\limpl}{$\supset$-L}
\newcommand{\ldimpl}{\supset\mbox{\rm -L}}
\newcommand{\limpr}{$\supset$-R}
\newcommand{\ldimpr}{\supset\mbox{\rm -R}}
\newcommand{\lnegl}{$\neg$-L}
\newcommand{\ldnegl}{\neg\mbox{\rm -L}}
\newcommand{\ldnegr}{\neg\mbox{\rm -R}}
\newcommand{\lalll}{$\forall$-L}
\newcommand{\ldalll}{\forall\mbox{\rm -L}}
\newcommand{\lallr}{$\forall$-R}
\newcommand{\ldallr}{\forall\mbox{\rm -R}}
\newcommand{\lsomel}{$\exists$-L}
\newcommand{\ldsomel}{\exists\mbox{\rm -L}}
\newcommand{\lsomer}{$\exists$-R}
\newcommand{\ldsomer}{\exists\mbox{\rm -R}}
\newcommand{\ldlamlr}{\lambda}
\newcommand{\sequent}[2]{\hbox{{$#1\ \longrightarrow\ #2$}}}
\newcommand{\prog}[2]{\hbox{{$#1\ \supset\ #2$}}}
\newcommand{\run}{\Gamma}

\newcommand{\Ibf}{{\bf I}}
\newcommand{\Cbf}{{\bf C}} 
\newcommand{\Cbfpr}{{\bf C'}}

\newcommand{\cprove}{\vdash_C}
\newcommand{\iprove}{\vdash_I}

\newsavebox{\lpartfig}
\newsavebox{\rpartfig}

% From the hohh section

\newenvironment{exmple}{
 \begingroup \begin{tabbing} \hspace{2em}\= \hspace{3em}\= \hspace{3em}\=
\hspace{3em}\= \hspace{3em}\= \hspace{3em}\= \kill}{
 \end{tabbing}\endgroup}
\newenvironment{example2}{
 \begingroup \begin{tabbing} \hspace{8em}\= \hspace{2em}\= \hspace{2em}\=
\hspace{10em}\= \hspace{2em}\= \hspace{2em}\= \hspace{2em}\= \kill}{
 \end{tabbing}\endgroup}

\newenvironment{example}{
\begingroup  \begin{tabbing} \hspace{2em}\= \hspace{3em}\= \hspace{3em}\=
\hspace{3em}\= \hspace{3em}\= \hspace{3em}\= \hspace{3em}\= \hspace{3em}\= 
\hspace{3em}\= \hspace{3em}\= \hspace{3em}\= \hspace{3em}\= \kill}{
 \end{tabbing} \endgroup }

\newcommand{\sand}{sand} % choice disjunction
\newcommand{\pand}{pand} % choice disjunction
\newcommand{\cor}{cor} % choice disjunction

\newcommand{\lb}{\langle}
\newcommand{\rb}{\rangle}
\newcommand{\pr}{prov}
\newcommand{\prG}{intp}
\newcommand{\prSG}{intp_E}
\newcommand{\intp}{intp_o}
\newcommand{\prove}{exec} % choice conjunction
\newcommand{\np}{invalid} % choice conjunction
\newcommand{\Ra}{\supset}  
\newcommand{\Cscr}{{\cal C}}
\newcommand{\seqweb}{SProlog}
\newcommand{\sprog}{{SProlog}}

\newtheorem{theorem}[lemma]{Theorem}

\newtheorem{proposition}[lemma]{Proposition}

\newtheorem{corollary}[lemma]{Corollary}
\newenvironment{proof}
     {\begin{trivlist}\item[]{\it Proof. }}%
     {\\* \hspace*{\fill} \end{trivlist}}

\newcommand{\seqand}{\prec}
\newcommand{\seqor}{\cup}
\newcommand{\seqandq}[2]{\prec_{#1}^{#2}}
\newcommand{\parandq}[2]{\land_{#1}^{#2}}
\newcommand{\exq}[2]{\exists_{#1}^{#2}}
\newcommand{\ext}{intp_G}
\newcommand{\etc}{{\em etc}}
\newcommand{\cf}{{\em c.f.}}

\begin{abstract}
This paper proposes a new approach to defining and expressing
 algorithms: the notion of {\it task logical} algorithms.  
This notion allows the user to  define an algorithm for a task $T$
 as  a  set of agents who can collectively perform $T$. 
This notion considerably simplifies the algorithm development process and can be seen as
 an integration of the sequential pseudocode and logical algorithms. 

This observation requires some changes to algorithm development process.
We propose a two-step approach: the first step is to define
an algorithm for a task $T$ via a set of agents that can collectively perform $T$.
The second step is to translate these agents
into  (higher-order) computability logic.

\end{abstract}

%% </local definitions here>

\section{Introduction}\label{sec:intro}

Traditional acquaintance with algorithm languages relates to the pseudocodes, also
known as imperative algorithms. Within this setting, algorithms are expressed as a
sequence of instructions.  Many algorithms in algorithm textbooks \cite{nea97} have been written
in pseudocodes. However, pseudocode is, in  a sense, a low-level language because the user must
specify the execution order.  In particular,  pseudocode is very awkward to use in expressing
nondeterministic algorithms such as graph problems where  execution orders are
typically unknown beforehand.

Logical algorithm language is a high-level language in which the execution order can be omitted.
Consequently, logical languages can express most deterministic/nondeterministic algorithms in a
concise way.  Traditional logical languages, however,  suffer from weak expressibility because
they are built around the notion of boolean logic (true/false) \cite{girard87tcs,hodas92ic}. 
It is possible to increase the
expressibility of logical languages by employing a task/game logic called computability logic (CL) \cite{Jap03, Jap08}, a  powerful logic which is built
around the notion of success/failure. The task logic offers many new, essential logical operators including
parallel conjunction/disjunction, sequential conjunction/disjunction, choice conjunction/disjunction,
\etc.

This paper proposes to use CL as an algorithm language. The distinguishing feature
of CL is that now each clause/agent is allowed to perform new, sophisticated tasks that have not
been supported by previous logical languages. While CL is an excellent algorithm language, it is
based on the first-order logic.  We also consider its higher-order extension where 
first-order
terms are replaced by higher-order terms. It is well-known that higher-order terms can describe
objects of function types including programs and formulas. Higher-order terms have proven
useful in many metalanguage applications such as theorem proving.

The remainder of this paper is structured as follows. We discuss a new way of defining
algorithms in the next section. In Section \ref{sec:modules}, we
present some examples.
Section~\ref{sec:conc} concludes the paper.

\section{Task Logical  Algorithms}\label{sec:logic}

 A {\it task logical} algorithm for a task $T$ is of the form

 \[ \sequent{c_1:T_1,\ldots,c_n:T_n}{d:T} \]
 
\noindent where $c_i:T_i$ represents  an agent $c_i$ who can do task $T_i$.
In the traditional developments of declarative algorithms, those $T_i$s are limited
to simple tasks such as computing recursive functions, relations or resources. 
Most complex
tasks such as interactive ones  are not permitted. 
In algorithm design, however, complex tasks are  desirable quite often. 
 Such examples include
  many OS processes, Web agents, \etc.

To define the class of computable tasks, we need a specification language. 
 An ideal  language would support an optimal
 translation of the tasks.
  We argue that a reasonable, high-level translation of the
tasks can be achieved via computability logic(CL)\cite{jap02,Jap03}. 
An advantage of CL over
other formalisms such as 
sequential pseudocode, linear logic\cite{girard87tcs}, \etc,  is that 
it can optimally encode a number of essential  tasks: nondeterminism, updates, \etc. Hence the main advantage of CL over 
other formalisms is the minimum (linear) size of the encoding.
 
We consider here a higher-order version of CL.
The logical language  we consider in this paper is built based on
   a typed lambda calculus.  Although types
 are strictly necessary, we will
omit these here because  their identity is not relevant in this paper.
An atomic formula is $(p\ t_1\ldots t_n)$
where $p$ is a (predicate) variable or non-logical constant and each $t_i$ is a lambda term. 

The basic operator in CL
is the reduction of the form $c: A \ra B$. This expression
means that the task  $B$ can be reduced to another task $B$.
The expression $c: A\land B$ means that the agent $c$ can perform  two tasks $A$ and $B$ in 
parallel.
The expression  $!A$ means that the agent  can perform the task $A$ repeatedly. 
The expression $c: A \adc B$ means  that the agent $c$ can perform  either task $A$ or $B$, 
regardless of what
the machine chooses.
The expression $c: \ada x A(x)$ means  that the agent $c$ can perform the task $A$, regardless of what
the machine chooses for $x$. The expression $c: A \add B$ means  that the agent $c$ can choose
and perform a true disjunct between $A$ and $B$.

The expression $c: \ade x A(x)$ means that the agent can choose a 
right value for $x$ so that it can perform the task $A$.
 We point the reader to \cite{Jap03,Jap08} to find out more about the
whole calculus of CL.

\section{Examples }\label{sec:modules}
% we need full linear logic + diffentia + proba
The notion of task logical algorithms makes algorithms simpler and versatile
compared to traditional approach. 
As an  example, we present the factorial
 algorithm to help understand this notion.  The factorial algorithm can be
defined using an agent $c$ whose tasks are 
 described below in English:

\begin{numberedlist}

\item   $c$ can either claim that $fact(0,1)$ holds, or

\item  can replace $fact(X,Y)$  by $fact(X+1,XY+Y)$.

\end{numberedlist}
\noindent
It is shown below that the above description can be translated into CL
formulas.
The following is a CL translation of the above algorithm,
where  the reusable action is preceded with $!$.
%We assume that $fact$ is a constant of type $int \ra int \ra o$.

\begin{exmple}
$c:(fact\ 0\ 1)\ \adc\  !\ \ada x\ada y\ ((fact\ x\ y)\ \ra\ (fact\ x+1\ xy+y))$.\\
\end{exmple}
\noindent A task is typically given by a user in the form of a query relative to
agents. Computation tries to solve the query with respect to the agent $c$. 
As an example, executing $\sequent{agent\ c}{\ada y \ade z fact(y,z)}$ would involve the
user choosing a value, say 5,  for $y$. This results
in the initial resource $fact(0,1)$ being transformed to $fact(1,1)$, then 
to $fact(2,2)$, and so on. It  will finally produce the desired
result $z = 120$  using the second conjucnt five times.

An example of interactive tasks is provided by the 
following  agent $t$ which has a lottery ticket. 
The ticket is represented as $ 0 \add\  1M$ which indicates that it
has two possible values, nothing or one million dollars.

 The following is a CL translation of the above algorithm.

\begin{exmple}
\>$t:   0 \add\  1M$.\\
\end{exmple}
\noindent  Now we want to execute $t$  to   obtain a final value. 
This interactive task is  represented by the query $t$.
 Now executing the program   
$\sequent{agent\ t}{agent\ t}$  would  produce the following question asked by the agent in the task of $ 0 \add\  1M$
 in the program: ``how much  is the final value?''. 
The user's response would be zero dollars. 
This move brings the task down to $\sequent{0}{agent\ t}$. Executing 
$\sequent{0}{agent\ t}$ would  require the 
machine to choose zero dollars in $ 0 \add\  1M$ for a success.

An example of parallel tasks is provided by the 
following two agents $c$ and $d$ working at a fastfood restaurant.  
The agent $c$ waits for a customer to pay money(at least three dollars), and
then generates a hamburger set consisting of a hamburger,
      a coke and a change. 
The agent $d$ waits for a customer to pay money(at least four dollars), and
then generates a fishburger set consisting of a fishburger,
      a coke and a change. 

The following is a CL translation of the above algorithm.

\begin{exmple}
\>$c: ! \ada x (\geq (x,3)\ \ra\ m(ham) \land  m(coke)\land m(x-3))$.\\
\>$d: ! \ada x (\geq (x,4)\ \ra m(fi) \land  m(coke)\land m(x-4))$.\\
\end{exmple}
\noindent  Now we want to execute $c$ and $d$ in parallel to   obtain a hamburger set and then
a fishburger set by interactively paying money to $c$ and $d$. 
This interactive task is  represented by the query $c\land d$.
 Now executing the program   
$\sequent{agent\ c, agent\ d}{agent\ c \land agent\ d}$  would  produce the following question asked by the agent in the task of 
$c$: ``how much do you want to pay me?''. The user's response would be five dollars. This move brings the task down to $m(ham) \land  m(coke)\land m(\$2)$ which would be a success. The task of $d$ would proceed similarly.

\newcommand{\prov}{pv}
\renewcommand{\ext}{pv_i}

As an example of higher-order algorithms, consider the interpreter for Horn clauses.
 It is described
by $G$- and $D$-formulas given by the syntax rules below:
\begin{exmple}
\>$G ::=$ \>   $A \sep  G\  and\  G \sep   some\  x\ G $ \\   \\
\>$D ::=$ \>  $A  \sep G\ imp\ A\ \sep all\ x\ D \sep  D\ and\ D$\\
\end{exmple}
\noindent
In the rules above,   
$A$  represents an atomic formula.
A $D$-formula  is called a  Horn
 clause. The expression $some\ x\ G$ involves bindings. We represent such
objects using lambda terms. For example, $all\ x\ p(x)$ is represented as
$all\ \lambda x (p\ x)$.
 
In the algorithm  to be considered, $G$-formulas will function as 
queries and $D$-formulas will constitute  a program. 

 We will  present an operational 
semantics for this language based on \cite{hol}. 
 Note that execution  alternates between 
two phases: the goal-reduction phase 
and the backchaining phase. Following Prolog's syntax, we assume that
names beginning with uppercase letters are quantified by $\ada$.

\begin{defn}\label{def:semantics}
Let $G$ be a goal and let $D$ be a program.
Then the notion of   executing $\lb D,G\rb$ -- $\prov\ D\ G$ -- 
 is defined as follows:
\begin{numberedlist}

\item  $bc\ D\ A\ A$. \% This is a success.

\item   $\prov\ D\ G_1\ \ra\ bc\ D\ (G_1\ imp\ A)\ A)$.

\item   $bc\ D\ (D\ X)\ A\  \ra\ bc\ D\ (all\ D)\ A$.

\item  $bc\ D\ D_1\ A  \lor
  bc\ D\ D_2\ A\ \ra\  bc\ D\ (D_1\ and\ D_2)\ A$.

\item   $atom\ A \land\ bc\ D\ D\  A\ \ra\ \prov\ D\ A$. \%  change to backchaining phase.
\item  $\prov\ D\ G_1  \land
  \prov\ D\ G_2\ \ra\ \prov\ D\ (G_1\ and\ G_2)$.

\item  $\prov\ D\ (G\ X)\ \ra\ \prov\ D\ (some\  G)$.

% This goal behaves as exclusive-OR. 

\end{numberedlist}
\end{defn}

\noindent  
In the rules (3) and (7), the symbol $X$  will be instantiated by a term.
In this context, consider the query $\prov\ (p\ a)\ (some\ \lambda x (p\ x))$.
In solving this query, $\prov\ (p\ a)\ (p\ a)$ will be formed and eventually 
solved.

The examples presented here have been of a simple nature. They are, however,
sufficient for appreciating the attractiveness of the algorithm development
process proposed here. We point the reader to \cite{MN87slp,lp2.7,hol} for
 more examples.

\section{Conclusion}\label{sec:conc}

A proposal for designing algorithms is given. It is based on the view that
 an algorithm for a task $T$ is a set of 
 agents who can collectively perform the task. The advantage of our
approach is that it simplifies the process of designing and writing algorithms
for most problems.

Our ultimate interest is in a procedure for carrying out computations
of the kind described above.  Hence it is important to realize this 
CL interpreter in an efficient way, taking advantages of  some techniques  discussed in 
\cite{ban02,CHP96,hodas92ic}. 

\section{Acknowledgements}

This paper was supported by Dong-A University Research Fund.

\bibliographystyle{ieicetr}% bib style

\begin{thebibliography}{1}

\bibitem{ban02}
M. Banbara.
\newblock {\em Design and implementation of linear logic programming languages}.
\newblock  Ph.D. Dissertation, Kobe University, 2002.

\bibitem{CHP96}
Iliano Cervesato, Joshua~S. Hodas, and Frank Pfenning.
\newblock Efficient resource management for linear logic proof search.
\newblock In {\em Proceedings of the 1996 Workshop on Extensions of Logic
  Programming}, LNAI 1050, pages 67 -- 81.


\bibitem{girard87tcs}
Jean-Yves Girard.
\newblock Linear logic.
\newblock {\em Theoretical Computer Science}, 50:1--102, 1987.

\bibitem{hodas92ic}
Joshus Hodas and Dale Miller.
\newblock Logic programming in a fragment of intuitionistic linear logic.
\newblock {\em Journal of Information and Computation}, 1994.
\newblock Invited to a special issue of submission to the 1991 LICS conference.

\bibitem{jap02}
G. Japaridze.
\newblock The logic of tasks.
\newblock {\em Annals of Pure and Applied Logic}, 117:263--295, 2002.

\bibitem{Jap03}
G. Japaridze.
\newblock Introduction to computability logic.
\newblock {\em Annals of Pure and Applied Logic}, 123:1--99, 2003.


\bibitem{Jap08}
G.~Japaridze.  
\newblock Sequential operators in computability logic.
\newblock {\em Information and Computation}, vol.206, No.12, pp.1443-1475, 2008.  

\bibitem{MN87slp}
D. Miller and G. Nadathur. 1987.
\newblock A logic programming approach to manipulating formulas and programs.
\newblock In {\em IEEE Symposium on Logic Programming}, {S.~Haridi}, Ed. IEEE
  Computer Society Press, 379--388.

\bibitem{lp2.7}
D. Miller and G. Nadathur. 1988.
\newblock $\lambda${P}rolog version 2.7.
\newblock Distributed in C-Prolog and Quintus Prolog source code.

\bibitem{hol}
D. Miller and G. Nadathur. 2012.
\newblock Programming with higher-order logic.
\newblock Cambridge University Press.

\bibitem{nea97}
R. Neapolitan and K. Naimipour.
\newblock {\em Foundations of Algorithms}.
\newblock Heath, Amsterdam, 1997.

\end{thebibliography}

\end{document}